\documentclass[preprint,12pt]{elsarticle}

\usepackage{longtable}
\usepackage{pdflscape} 
\usepackage{subfig}
\usepackage{amsmath}

\journal{Journal of \LaTeX\ Templates}

\begin{document}

\begin{frontmatter}






\title{Systematic Analysis of Double-Beta Decay Half Lives}
\cortext[cor1]{Corresponding author}
\author[label1]{B. Pritychenko\corref{cor1}}
\ead{pritychenko@bnl.gov}
\address[label1]{National Nuclear Data Center, Brookhaven National Laboratory, \\ Upton, NY 11973-5000, USA}

\begin{abstract}
Evaluated $2\beta^{-}(2\nu)$ half-lives and their systematics were reexamined in the framework of a phenomenological approach. Decay rate dependence on nuclear deformation, decay energy, shape coexistence, and forbidden transitions  was observed. The following analysis showed distinct impacts of decay energy on half lives, and deformation parameters on effective nuclear matrix elements. These findings were used to predict $T_{1/2}$ for 36 isotopes of interest. Present work results were compared with published data.
\end{abstract}


\end{frontmatter}


\section{Introduction}

Double-beta decay was introduced by M. Goeppert-Mayer in 1935 \cite{35Go} as a nuclear disintegration with simultaneous emission of  two electrons and two neutrinos
\begin{equation}
\label{myeq.2b}  
(Z,A) \rightarrow (Z+2,A) + 2 e^{-} + 2\bar{\nu}_{e}
\end{equation}

 There are several  double-beta decay processes: $2 \beta^{-}$,  $2 \beta^{+}$, $\epsilon\beta^{+}$, 2$\epsilon$ and two major decay modes: 
 two-neutrino (2$\nu$) and neutrinoless (0$\nu$).  
   2$\nu$-mode is not prohibited by conservation laws and occurs as a second-order process compared to the regular $\beta$-decay.  0$\nu$-mode varies from the 2$\nu$-mode by the fact that no neutrinos are emitted during the decay. This normally requires that the lepton number is not conserved and neutrino should contain a small fraction of massive particles that equals to its anti-particle (Majorana neutrino).  
   
\begin{figure}
\centering
\includegraphics[width=0.8\textwidth]{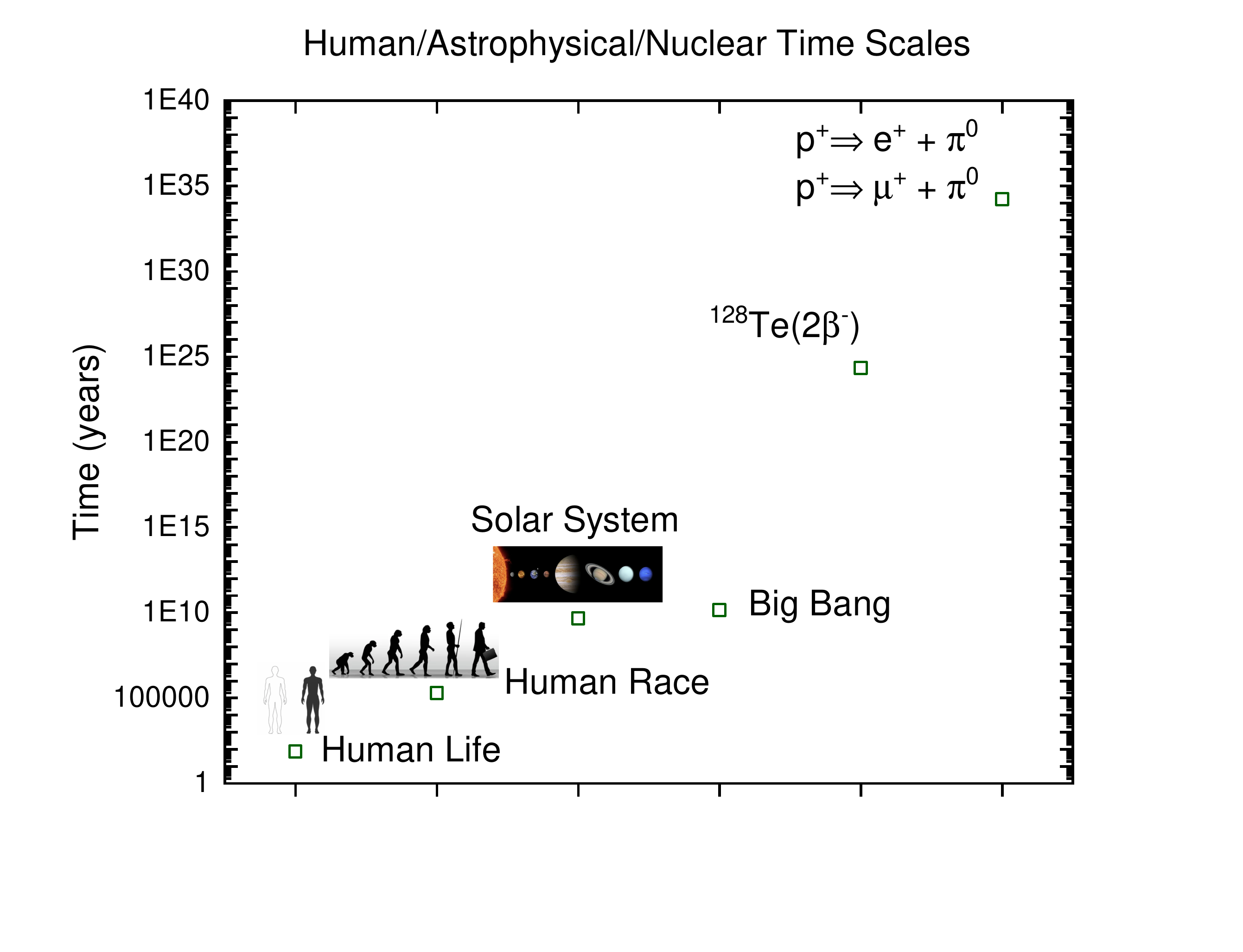}
\caption{Human/Astrophysical/Nuclear Time Scales. The nuclear time scale is based on the $^{128}$Te(2$\beta^{-}$) half-life and the lower limit for hypothetical proton decay.}
\label{fig:Time}
\end{figure}

Two-neutrino decay mode is observed in more than a dozen nuclei~\cite{Prit14,02Tr,Bar20}, and experimental half-lives exceed the age of the Universe~\cite{Goh21} by many orders of magnitude.  2$\beta (2\nu)$-decay is the rarest observed nuclear decay in nature. Figure~\ref{fig:Time} data show that double-beta decay half-lives exceed human and cosmological lifetimes, and only the hypothetical decay of protons into other subatomic particles~\cite{Sak67} is rarer. 
   
Since 1935, thousands of double-beta decay works were published in the literature. Analysis of the Nuclear Science References database~\cite{Pr11}  shows that $\approx 35 \%$ of papers are experimental, and the rest are theoretical. Published theoretical calculations provide a very extensive range of possible scenarios and observables. Due to the lack of a nuclear theory that comprehensively describes atomic nuclei from calcium to uranium with good precision, nuclear physicists often use different models to calculate the properties of chosen nuclei. To resolve variations in published predictions and produce realistic recommendations for the sensitivity estimates of future experiments across the nuclear chart, it is worth to analyze the evaluated half-lives, deduce trends, and constrain the theoretical pursuits using a phenomenological approach~\cite{Gro62,Prit23}.

\section{Compilation and Evaluation of Experimental Data}
Double-beta decay is an important nuclear physics phenomenon and experimental results in this field have been compiled by several groups~\cite{02Tr,Prt06,Work22}, and multiple evaluations have been produced~\cite{Prit14,El02,Bar20}. In this work, we will select the NNDC evaluation~\cite{Prit14} that was produced using the internationally recommended nuclear structure and decay evaluation practices~\cite{ensdf,Dim20,NSDD}. 

\begin{table*}[!htb]
\centering
\caption{$2 \beta^{-}$ decay $Q$-values~\cite{Wang21}, deformation parameters~\cite{Prit16},  evaluated  {\it T}$_{1/2}$(2$\beta$) and effective nuclear matrix elements~\cite{Prit14}.}
\begin{tabular}{c|c|c|c|c}
\hline
\hline
Nucleus  & $Q_{2\beta}$-value (MeV) & $\beta_2$ & $T_{1/2}^{2 \nu, eval.}$(y) & $M_{eff}^{2\nu, eval.}$  \\
\hline

$^{48}$Ca & 4.26808 & 0.1054(50) & (4.39$\pm$0.58)x10$^{19}$ & 0.0383$\pm$0.0025 \\ 		
$^{76}$Ge & 2.03906 & 0.2650(15) & (1.43$\pm$0.53)x10$^{21}$ & 0.120$\pm$0.021 \\ 		
$^{82}$Se &   2.9979  & 0.1939(53) & (9.19$\pm$0.76)x10$^{19}$ & 0.0826$\pm$0.0034\\ 		
$^{96}$Zr &  3.35603  &  	0.0615(33) & (2.16$\pm$0.26)x10$^{19}$ & 0.0824$\pm$0.0050 \\ 		
$^{100}$Mo &  3.03436  &  0.2340(49) & (6.98$\pm$0.44)x10$^{18}$ &   0.208$\pm$0.007 \\ 			
$^{116}$Cd & 2.81349  & 0.194(44) & (2.89$\pm$0.25)x10$^{19}$ &  0.112$\pm$0.005 \\ 		
$^{128}$Te & 0.8667  &  0.1862(37) & (3.49$\pm$1.99)x10$^{24}$  & 0.0326$\pm$0.0093 \\
$^{130}$Te &  2.52751  & 0.1185(20) & (7.14$\pm$1.04)x10$^{20}$ &  0.0303$\pm$0.0022 \\ 	
$^{136}$Xe &  2.45791  & 0.0949(75)& (2.34$\pm$0.13)x10$^{21}$ &  0.0173$\pm$0.0005  \\	
$^{150}$Nd &   3.37138  & 0.2825(16) & (8.37$\pm$0.45)x10$^{18}$ & 0.0572$\pm$0.0015 \\ 		
$^{238}$U  & 1.1446  & 0.2741(36) & (2.00$\pm$0.60)x10$^{21}$ & 0.185$\pm$0.028 \\
\hline
\hline
\end{tabular}
\label{table1}
\end{table*}

\section{Analysis of Evaluated Half Lives}

Table \ref{table1} shows the NNDC evaluated values which were deduced using the limitation of minimum statistical weight procedures~\cite{98Lwe}. All final results from distinct experiments were included in the evaluation process.   It is helpful to analyze the evaluated half-lives using the Grodzins' method \cite{Gro62,Prit23}.  
In the analysis, we will consider only  $2 \beta^{-}$-decay   0$^{+}$ $\rightarrow$ 0$^{+}$ ground-state transitions, i.e. transitions 
without $\gamma$-rays.  $2 \beta^{-}$-decay relatively high $Q$-values~\cite{Wang21} and deformation parameters ($\beta_{2}$)~\cite{Prit16} lead the decay to the level of sensitivity of modern experiments, and it has been observed in 11 nuclei. The eleven parent nuclei from $Z$=20 to $Z$=92 provide a reasonable statistical sample for investigating systematic trends. $T_{1/2}^{2\nu}$ values are often described as follows \cite{87Bo}
\begin{equation}
\label{myeq.Half}  
(T_{1/2}^{2\nu} (0^{+} \rightarrow 0^{+}))^{-1} = G^{2 \nu} (E,Z) \times |M^{2 \nu}_{GT} - \frac{g^{2}_{V}}{g^{2}_{A}} M^{2 \nu}_{F}|^{2}, 
\end{equation}
where the function  $G^{2 \nu} (E,Z)$ results from lepton phase space integration and contains all relevant constants, and $ |M^{2 \nu}_{GT} - \frac{g^{2}_{V}}{g^{2}_{A}} M^{2 \nu}_{F}|$ is nuclear matrix element. From the Eq. \ref{myeq.Half} one may conclude that decay half-lives depend on decay energy, nuclear charge, and deformation.

Previously, Primakoff and Rosen \cite{59Pr,69Pr} predicted that for $2\beta$(2$\nu$) decay, the phase space available to the (four) emitted leptons is roughly proportional to the 7$^{th}$ through 11$^{th}$ power of energy release, and $2\beta(2\nu)$ transition probability, $W(2\nu)$, depends on  the maximum kinetic energy of the electron in the units of electron mass as 

\begin{equation}
\label{myeq.PR}  
W(2\nu) \sim T^{7}_{0}[1 + \frac{T_{0}}{2} + \frac{T_{0}^2}{9} + \frac{T_{0}^3}{90} + \frac{T_{0}^4}{1980} ],
\end{equation}
where $T_{0}$=$Q_{2\beta}/mc^{2}$, $Q_{2\beta^-}$ = $M(A,Z) - M(A,Z+2)$, and $M(A,Z)$ is the atomic mass of the isotope with mass number $A$ and atomic number $Z$~\cite{87Bo,13Sa}. Furthermore,  it is known half-lives depend on dimensionless Coulomb energy parameter \mbox{$\xi$ $\approx$ $Z A^{-1/3}$} \cite{98Bo,83Ab,13Sa}, and  
several authors showed the relation between nuclear deformation~\cite{Lob70} and the 2$\nu$ mode nuclear transition matrix element~\cite{Vog85,Trol96,Singh07}. 

These findings and formula~\ref{myeq.Half}  suggest a possibility for the phenomenological description of $2 \beta$-transition half lives as~\cite{Prit10}
\begin{equation}
\label{myeq.Fit}  
T_{1/2}^{2\nu} (0^{+} \rightarrow 0^{+}) \approx \frac{x}{\xi^{2} Q_{2\beta}^{y} \beta_{2}^{z}},
\end{equation}
where $T_{1/2}^{2\nu}$ in years, $Q_{2\beta}$ in MeV and  $x$, $y$ and $z$ are fitting parameters.  The fitting parameters for evaluated half-lives~\cite{Prit14} are given in Table~\ref{FitParam}. The least-squares fit data confirm theoretical findings that half-lives are inversely proportional to decay $Q$-value (transition energy) and quadrupole deformation parameter values in eight and fifth degrees~\cite{69Pr,Vog85}, respectively. The observed transition energy dependence validates the previous result on $^{128,130}$Te half-life systematic~\cite{Prit14,Prit10}, and is consistent with the prediction of Primakoff and Rosen~\cite{59Pr,69Pr}.
\begin{table}
\caption{\label{FitParam} Fitting parameters $x$, $y$ and $z$ numerical values.}
\begin{center}
\begin{tabular}{c|c|c}
\hline \hline
$x$ & $y$ &  $z$ \\ 
\hline
0.00301$\times 10^{24}$ &	 8.16(59)  & 4.98(32)	 \\
\hline \hline
\end{tabular}
\end{center}
\end{table}

\section{Phenomenological Approach Predictions}

\begin{figure}[ht]
\centering
\includegraphics[width=0.8\textwidth]{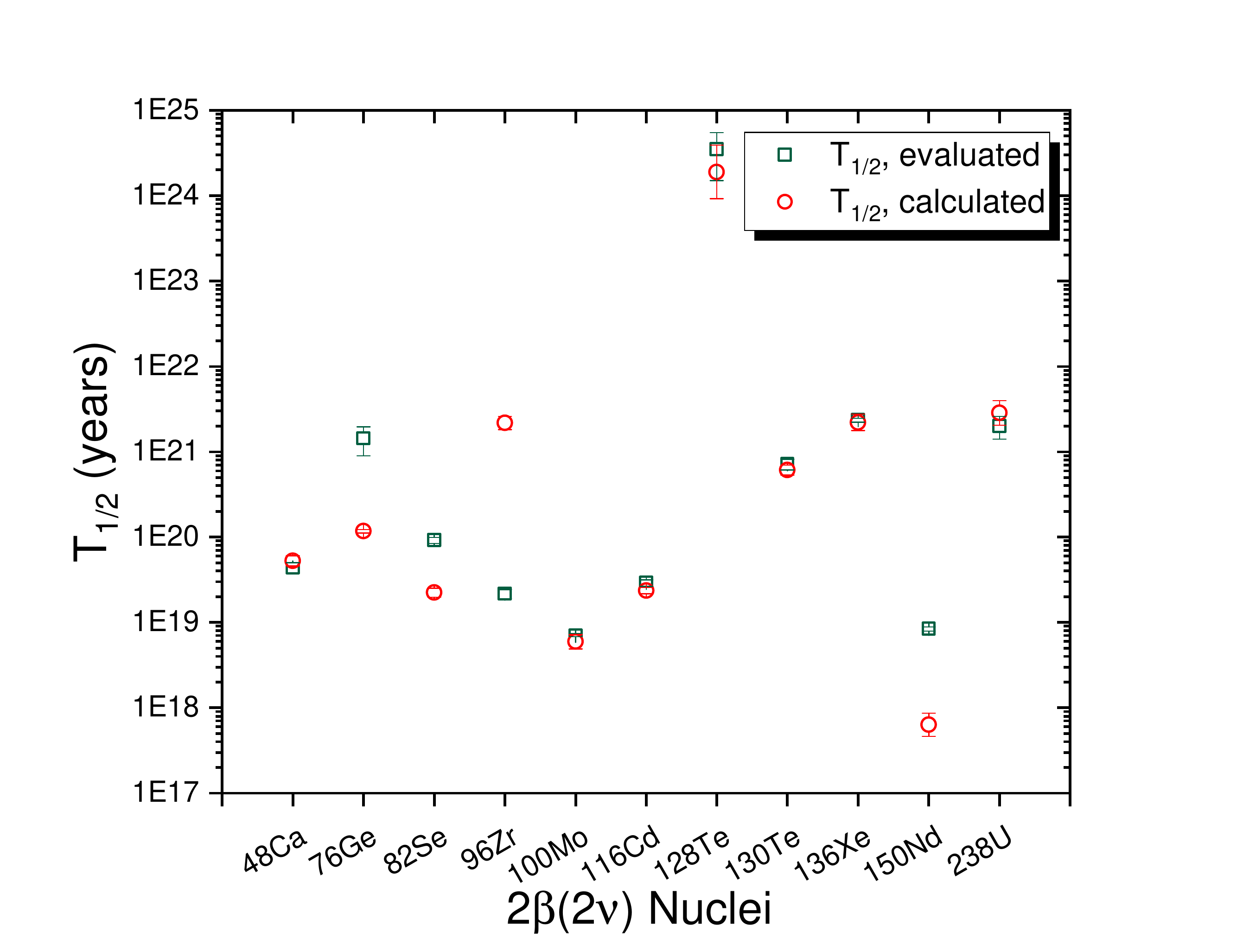}
\vspace{-0.5cm}
\includegraphics[width=0.8\textwidth]{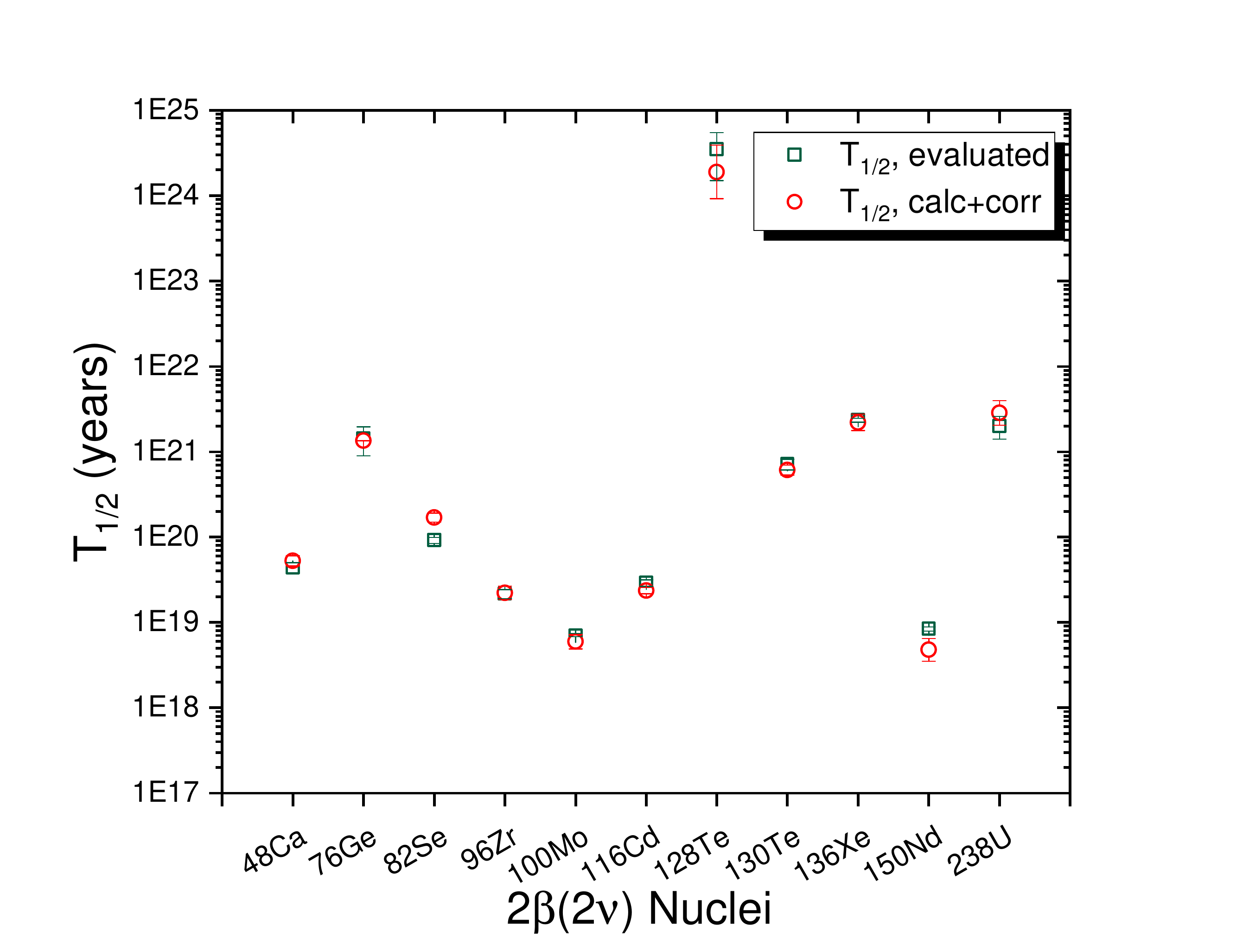}
\caption{Calculated (fitted) and evaluated double beta decay half-lives. The original and corrected calculated values are shown in the upper and lower panels, respectively.}
\label{fig:comp2}`
\end{figure}

Eq. \ref{myeq.Fit} allows us to calculate half-lives for all nuclei of interest as shown in Table \ref{table2}. The comparison of the calculated and evaluated half-lives is shown in Fig.~\ref{fig:comp2}. Further analysis of plotted data reveals one case of overprediction in $^{96}$Zr, the three cases of the current fit under prediction: $^{76}$Ge, $^{82}$Se, and $^{150}$Nd that are considered below.

The tiny $^{96}$Zr deformation parameter of 0.0615(33)~\cite{Prit16} is responsible for very large predicted half-life. It is lower than the semi-magic $^{90}$Zr parameter of 0.0907(24). Both values have been obtained from the quadrupole collectivities of the first 2$^{+}_{1}$ states in zirconium nuclei. The recent review of the first excited states spin and parity assignments in even-even nuclei shows that 2$^{+}_{1}$ is observed in 96$\%$ of cases~\cite{Prit22}, and 0$^{+}_{1}$, 1$^{-}_{1}$, and 3$^{-}_{1}$ assignments are also present. The 0$^{+}_{1}$ first excited state is present in $^{96}$Zr, and the state indicates shape coexistence when the deformed 0$^{+}_{1}$ and spherical 2$^{+}_{1}$ excited states coexist in the same nucleus. The semi-empirical formula requires an increased deformation parameter $\beta_2 (increased) \approx  \frac{5}{2} \beta_2$ to reproduce the evaluated half-life. The 0$^{+}_{1}$ first excited states are present in 3$\%$ of even-even nuclei, and the correction of 2$\nu$-mode of 2$\beta$-decay is limited to $^{96}$Zr and $^{98}$Mo.

The detailed analysis of nuclear properties of even-even atomic nuclei reveals that in $^{76}$Ge$\rightarrow ^{76}$Se, $^{82}$Se$\rightarrow ^{82}$Kr, and $^{150}$Nd$\rightarrow ^{150}$Sm transitions the intermediate nuclei ($^{76}$As,$^{82}$Br, and $^{150}$Pm) ground-state parity is negative while in all other cases, it is positive. 
The negative parity suggests that these transitions are more complex than the Fermi/Gamow-Teller allowed and superallowed transitions where the selection rules for beta decay involve no parity change of the nuclear state (The spin of the parent nucleus can either remain unchanged or change by $\pm$1), and first forbidden transitions could be present. Following the $^{96}$Zr recipe, a reduced deformation parameter $\beta_2 (reduced)  \approx \frac{2}{3} \beta_2$ was explored for explanation of the predicted values.   The subsequent analysis of the shown in Table~\ref{table2} data reveals that simple deformation parameter adjustment improvements are not sufficient for resolving all discrepancies, and more sophisticated assumptions are needed.

\begin{figure}[ht]
 \centering
\includegraphics[width=0.8\textwidth]{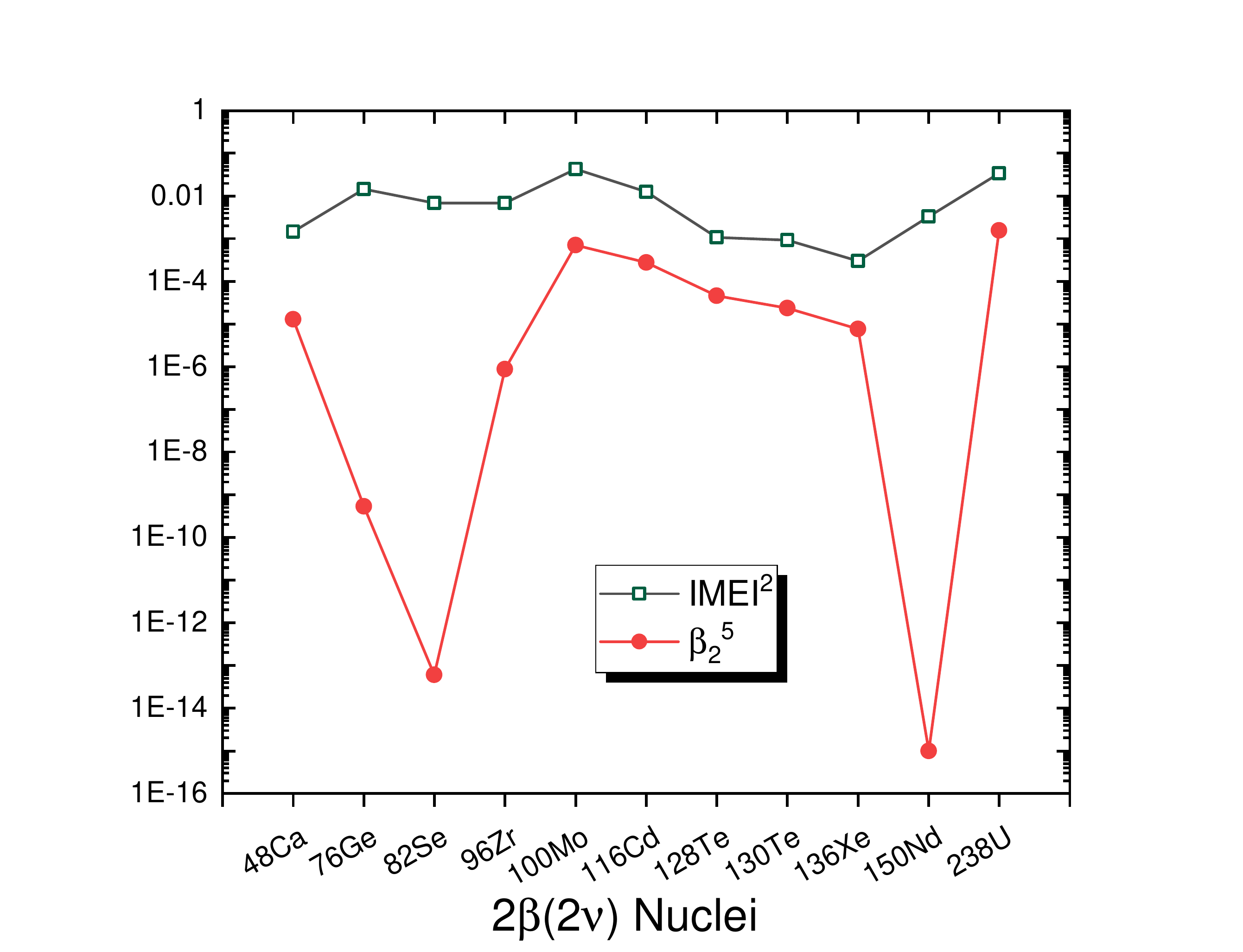}
\caption{Correlations between $\beta_2^5$ and $|M_{eff}^{2\nu}|^2$ in $2 \beta^{-}$ nuclei of interest.}
\label{fig:corr}
\end{figure}

Formula~\ref{myeq.Half}  indicates that half-lives are defined by phase factors and nuclear matrix elements, and the impact of decay energies (phase factors) on half-lives is visible in tellurium nuclei. The P. Vogel's analysis~\cite{Vog85} suggests direct correlations between nuclear deformation parameters and double-beta decay nuclear matrix elements in parent nuclei. Therefore, we would explore these correlations using the Table~\ref{FitParam} values. Figure~\ref{fig:corr} shows values of the fifth power of quadrupole deformation parameters ($\beta_2^5$)~\cite{Prit16} and squared effective nuclear matrix elements ($|M_{eff}^{2\nu}|^2$)~\cite{Prit14} for nuclei of interest. The plotted data imply that the above-mentioned quantities are consistent with the theoretical predictions~\cite{Vog85,Trol96,Singh07}, they are mutually related and follow the previously observed trends and deviations ($^{76}$Ge,$^{82}$Se, $^{96}$Zr, and $^{150}$Nd). Therefore, it is possible to express them as
\begin{equation}
\label{myeq.Corr}  
 |M_{eff}^{2\nu}|^2 \approx k \times \beta_2^5,
\end{equation}
where $k\sim$29.3 is obtained using the complementary fitting procedures.   
The mutual correlations between deformation parameters and effective nuclear matrix elements provide a clear validation of the double-beta decay formalism~\ref{myeq.Half} and bring credibility to the calculated half-lives.

The present work examples explain the issues with the predicted half-life fits and show the impact of nuclear structure effects on their values. The phenomenological predictions capitalize on the previous effort when a linear dependence between deformation parameters and nuclear matrix elements was explored in 2010~\cite{Prit10}.  A similar task was recently performed by a Nanjing-Lanzhou group using the  Geiger-Nuttall law and the Viola-Seaborg formula arguments~\cite{Ren14}. The Nanjing-Lanzhou group assumed that the logarithm of half-life is inversely proportional to decay energy as a starting point and deduced fitting parameters to reproduce the Institute of Theoretical and Experimental Physics (ITEP), Moscow values~\cite{Bar10}. Unfortunately, heavy reliance on $\alpha$- and $\beta$-decay fitting procedures and the addition of a $S=2$ parameter for neutron number magic nuclei are not sufficient for the description of the large variety of nuclear structure effects in 2$\beta^{-}$-decay nuclei and reliable predictions.
\begin{landscape}
\begin{longtable}{c|cccccc}

\caption{Phenomenological Rule Predictions: $2 \beta^{-}$(2$\nu$)-decay calculated, evaluated and experimental values for  0$^{+}$ $\rightarrow$ 0$^{+}$ transitions. 
Nuclei and half-lives with probable contributions of intermediate nucleus negative parity states,  shape coexistence,  and  $0 \nu$ decay mode are marked 
with $\dagger$, $\ddagger$, and  * characters, respectively. Calculated half-lives corrected for negative parity and shape coexistence effects using factors of 7.6 and  0.01024, respectively, are shown in the fifth column. 
 } \\
\hline \hline
  \textbf{Nuclide}  &  $Q_{2\beta}$ (MeV)~\cite{Wang21} & $\beta_{2}$~\cite{Prit16} & $T_{1/2}^{2 \nu, calculated}$(y) &  $T_{1/2}^{2 \nu, calc+corr}$(y) & $T_{1/2}^{2 \nu, eval.}$(y)~\cite{Prit14}  &  $T_{1/2}^{2 \nu , exp.}$(y)~\cite{Prt06}  \\
\hline
\endfirsthead
\hline
  \textbf{Nuclide}  &  $Q_{2\beta}$ (MeV)~\cite{Wang21}  & $\beta_{2}$~\cite{Prit16} & $T_{1/2}^{2 \nu, calculated}$(y) &  $T_{1/2}^{2 \nu, calc+corr}$(y) & $T_{1/2}^{2 \nu, eval.}$(y)~\cite{Prit14}  &  $T_{1/2}^{2 \nu , exp.}$(y)~\cite{Prt06}  \\
\hline 
\endhead
\hline
\multicolumn{6}{r}{\textit{Continued on next page ...}} \\
\endfoot
\hline
\endlastfoot  

$^{46}$Ca  &	0.9887 	&	0.1468 &	(1.496$^{1.287}_{69.19}$)x10$^{24}$ &	&                              & \\
$^{48}$Ca  &	4.26808 &	0.1054 &	(5.259$^{0.767}_{0.670}$)x10$^{19}$ &	&  (4.39$\pm$0.58)x10$^{19}$   &  \\
$^{70}$Zn  &	0.9973 	&	0.2218 &	(1.050$^{0.653}_{0.403}$)x10$^{23}$ &	&                              & $>$3.8$\times$10$^{18}$ \\
$^{76}$Ge$\dagger$  &	2.03906 &	0.265  &	(1.173$^{0.0054}_{0.0054}$)x10$^{20}$ & (1.347$^{0.004}_{0.004}$)x10$^{21}$   &  (1.43$\pm$0.53)x10$^{21}$   &   \\
$^{80}$Se  &	0.134 	&	0.2314 &	(9.389$^{39.71}_{7.594}$)x10$^{29}$  &   &        					   & $>$3.5$\times$10$^{13}$* \\
$^{82}$Se$\dagger$   &	2.9979 	&	0.1939 &	(2.231$^{0.29}_{0.26}$)x10$^{19}$  &  (16.955$^{2.204}_{1.976}$)x10$^{19}$    &  (9.19$\pm$0.76)x10$^{19}$   &  \\
$^{86}$Kr  &	1.25742 &	0.1347 &	(1.521$^{0.997}_{0.601}$)x10$^{23}$ &	&                              &  $>$2.0$\times$10$^{8}$* \\
$^{94}$Zr  &	1.14475 &	0.0882 &	(2.303$^{2.332}_{1.156}$)x10$^{24}$ &    &                              & $>$5.2$\times$10$^{19}$*  \\
$^{96}$Zr$\ddagger$  &	3.35603 &	0.0615 &	(2.171$^{0.432}_{0.354}$)x10$^{21}$  & (2.223$^{0.442}_{0.362}$)x10$^{19}$   &  (2.16$\pm$0.26)x10$^{19}$   &  \\
$^{98}$Mo$\ddagger$  &	0.109 	&	0.1692 &	(1.806$^{9.985}_{1.529}$)x10$^{31}$ & (1.850$^{10.224}_{1.566}$)x10$^{29}$   &                              & $>$5.0$\times$10$^{13}$*  \\
$^{100}$Mo &	3.03436 &	0.234  &	(5.930$^{1.242}_{1.027}$)x10$^{18}$ &    &  (6.98$\pm$0.44)x10$^{18}$   &  \\
$^{104}$Ru &	1.2994 	&	0.2717 &	(2.670$^{0.801}_{0.616}$)x10$^{21}$  &   &                              & $>$1.9$\times$10$^{20}$*  \\
$^{110}$Pd &	2.0171 	&	0.2562 &	(9.390$^{0.207}_{0.202}$)x10$^{19}$ &    &                              &  $>$6.0$\times$10$^{16}$ \\
$^{114}$Cd &	0.5448 	&	0.1888 &	(1.759$^{2.532}_{1.038}$)x10$^{25}$ &    &                              & $>$1.3$\times$10$^{18}$ \\
$^{116}$Cd &	2.81349 &	0.194  &	(2.362$^{0.211}_{0.194}$)x10$^{19}$  &   &  (2.89$\pm$0.25)x10$^{19}$   &  \\
$^{122}$Sn$\dagger$ &	0.3733 	&	0.1027 &	(7.685$^{0.208}_{5.611}$)x10$^{27}$ &  (5.841$^{0.158}_{4.264}$)x10$^{28}$  &                              & $>$1.3$\times$10$^{13}$* \\
$^{124}$Sn$\dagger$ &	2.2927 	&	0.0942 &	(4.416$^{1.348}_{1.033}$)x10$^{21}$ &  (3.356$^{1.025}_{0.785}$)x10$^{22}$  &                              & $>$1.0$\times$10$^{17}$* \\
$^{128}$Te &	0.8667 	&	0.1358 &	(1.890$^{2.006}_{0.973}$)x10$^{24}$  &   &  (3.49$\pm$1.99)x10$^{24}$*  &  \\
$^{130}$Te &	2.52751 &	0.1185 &	(6.064$^{0.880}_{0.768}$)x10$^{20}$  &   &  (7.14$\pm$1.04)x10$^{20}$   &  \\
$^{134}$Xe &	0.82417 &	0.1158 &	(6.024$^{7.346}_{3.328}$)x10$^{24}$  &   &                              & $>$8.7$\times$10$^{20}$ \\
$^{136}$Xe &	2.45791 &	0.0949 &	(2.200$^{0.550}_{0.440}$)x10$^{21}$  &   &  (2.34$\pm$0.13)x10$^{21}$   &  \\
$^{142}$Ce$\dagger$ &	1.4172 	&	0.1245 &	(4.539$^{2.658}_{1.676}$)x10$^{22}$ &  (3.450$^{2.020}_{1.274}$)x10$^{23}$  &                              & $>$1.4$\times$10$^{18}$ \\
$^{146}$Nd$\dagger$ &	0.0704 	&	0.1512 &	(7.158$^{0.555}_{6.341}$)x10$^{32}$ &  (5.440$^{0.422}_{4.820}$)x10$^{33}$   &                              & $>$3.0$\times$10$^{9}$* \\
$^{148}$Nd$\dagger$ &	1.928 	&	0.2004 &	(3.305$^{0.448}_{0.394}$)x10$^{20}$ &  	(2.512$^{0.340}_{0.300}$)x10$^{21}$  &                              &  \\
$^{150}$Nd$\dagger$  &	3.37138 &	0.2825 &	(6.310$^{2.315}_{1.694}$)x10$^{17}$ &  (4.795$^{1.696}_{1.288}$)x10$^{18}$   &  (8.37$\pm$0.45)x10$^{18}$   &  \\
$^{154}$Sm$\dagger$ &	1.2508 	&	0.3404 &	(7.758$^{1.840}_{1.487}$)x10$^{20}$ &  	(5.896$^{1.398}_{1.130}$)x10$^{21}$   &                              &  \\
$^{160}$Gd$\dagger$ &	1.7304 	&	0.3511 &	(4.531$^{0.052}_{0.051}$)x10$^{19}$  &  (3.444$^{0.040}_{0.038}$)x10$^{20}$   &                              & $>$1.9$\times$10$^{19}$ \\
$^{170}$Er$\dagger$ &	0.6564 	&	0.3368 &	(1.400$^{1.142}_{0.629}$)x10$^{23}$ &  (1.064$^{0.868}_{0.478}$)x10$^{24}$   &                              &  \\
$^{176}$Yb &	1.0851 	&	0.3014 &	(3.889$^{1.551}_{1.109}$)x10$^{21}$ &    &                              &  \\
$^{186}$W$\dagger$  &	0.4914 	&	0.2257 &	(9.782$^{14.17}_{5.787}$)x10$^{24}$ & (7.434$^{10.769}_{4.398}$)x10$^{25}$  &                              & $>$2.3$\times$10$^{19}$ \\
$^{192}$Os &	0.406 	&	0.1639 &	(2.213$^{4.506}_{1.484}$)x10$^{26}$  &   &                              & $>$2.0$\times$10$^{20}$* \\
$^{198}$Pt$\dagger$ &	1.0503 	&	0.1137 &	(5.674$^{5.379}_{2.761}$)x10$^{23}$ &  	(4.312$^{4.088}_{2.098}$)x10$^{24}$   &                              &  \\
$^{204}$Hg$\dagger\ddagger$ &	0.4197 	&	0.0683 &	(1.240$^{3.646}_{0.926}$)x10$^{28}$ &  (9.650$^{28.370}_{7.200}$)x10$^{26}$  &                              & $>$8.8$\times$10$^{8}$ \\
$^{232}$Th$\dagger$ &	0.8373 	&	0.2571 &	(5.177$^{3.701}_{2.158}$)x10$^{22}$ &  (3.935$^{2.813}_{1.640}$)x10$^{23}$  &                              & $>$2.1$\times$10$^{9}$ \\
$^{238}$U  &	1.1446 	&	0.2741 &	(2.857$^{1.135}_{0.812}$)x10$^{21}$  &   &  (2.00$\pm$0.60)x10$^{21}$*  &  \\
$^{244}$Pu$\dagger$ &    1.3542 	&   0.2954 &    (5.115$^{1.224}_{0.988}$)x10$^{20}$  &  (3.887$^{0.930}_{0.751}$)x10$^{21}$   &                              &     $>$1.1$\times$10$^{18}$      \\
\label{table2}
\end{longtable}
\end{landscape}

\section{Conclusion}
The two-neutrino mode of 2$\beta^{-}$-decay has been observed by various groups in 11 even-even nuclei~\cite{02Tr,Prt06,ensdf}, and several sets of evaluated half-lives were deduced~\cite{Prit14,Bar20,ensdf}.  $^{128,130}$Te data analysis led to observation of $\frac{1}{E^{8}}$ energy trend for  $T_{1/2}^{2 \nu}$  values~\cite{Prit14}. The addition of 
quadrupole deformation parameters, for simulation of the nuclear structure effects in predicted half-lives, revealed the effects of first-forbidden transitions and shape coexistence in several nuclei.    The subsequent analysis showed the independent impact of deformation parameters on effective nuclear matrix elements. 
The decay energy and quadrupole deformation parameter trends were used to explain nuclear systematics 
and calculate 2$\beta^{-}$(2$\nu$)-decay half-lives for all nuclei of interest.

\section{Acknowledgments}
\label{sec:Acknowledgements}

The author is thankful to David Brown (Brookhaven National Laboratory), Mihai Horoi (Central Michigan University), Remco Zegers (Michigan State University)  for helpful discussions, and the referee for very constructive comments that helped to improve the manuscript. 
Work at Brookhaven was funded by the Office of Nuclear Physics, Office of Science of the U.S. Department
of Energy, under Contract No. DE-SC0012704 with Brookhaven Science Associates, LLC.  \\ \\



\end{document}